\RequirePackage{fix-cm}
\documentclass[twocolumn]{svjour3}       
\smartqed  
\usepackage[pdftex]{graphicx}
\usepackage[left,columnwise]{lineno}
\usepackage{latexsym}
\usepackage{graphicx}
\usepackage{epsfig}
\usepackage{xcolor}
\usepackage{amssymb}
\linenumbers
\begin{document}

\title{One-Loop Correction to The Casimir Energy in Lifshitz-Like Theory 
}


\author{M A Valuyan}
\authorrunning{M A Valuyan} 
\date{\today}
\institute{M.A. Valuyan \at
              Department of Physics, Semnan Branch, Islamic Azad University, Semnan, Iran \\
              Energy and Sustainable Development Research Center, Semnan Branch, Islamic Azad University, Semnan, Iran\\
              Tel.: +98-23-33654040\\
              Fax: +98-23-33654036\\
              \email{m-valuyan@sbu.ac.ir; m.valuyan@semnaniau.ac.ir}}
\maketitle
\begin{abstract}
In the present article, Radiative Correction (RC) to the Casimir energy was computed for the self-interacting massive / massless Lifshitz-like scalar field, confined between a pair of plates with Dirichlet and Mixed boundary conditions in $3+1$ dimensions. Moreover, using the results obtained for the Dirichlet Casimir energy, the RC to the Casimir energy for Periodic and Neumann boundary conditions were also draw outed. To renormalize the bare parameters of the Lagrangian, a systematic perturbation expansion was used in which the counterterms were automatically obtained in a position - dependent manner. In our view, the position dependency of the counterterm was allowed, since it reflected the effects of the boundary condition imposed or the background space in the problem. All the answers obtained for the Casimir energy were consistent with well-known physical expects. In a language of graphs, the Casimir energy for the massive Lifshitz-like scalar field confined with four boundary conditions (Dirichlet, Neumann, Mixed, and Periodic) was also compared to each other, and as a concluding remark, the sign and magnitude of their values were discussed.
\end{abstract}
\vspace{-0.5cm}
\section{Introduction}\label{sec: introduction}
Nowadays, the Casimir effect is known as one of the most notable consequences of vacuum quantum fluctuations. This effect was first theoretically predicted by H.B.G. Casimir in 1948\,\cite{Casimir}, and experimentally confirmed 10 years later by Sparnaay \cite{sparnaay}. In addition to the interest in the Casimir energy for free quantum fields, the Radiative Correction\,(RC) to the Casimir energy was investigated, and this category of the problem was developed for various self-interacting quantum fields under different boundary conditions in multiple geometries \cite{Bordag.et.al.1,Bordag.et.al.2,Dowker.3,All.Casimir.litertaure.1,Milton.book.,All.Casimir.litertaure.2,Curved.1,Curved.2,All.Casimir.litertaure.3,RC.1,RC.3,All.Casimir.litertaure.4,All.Casimir.litertaure.5,RC.4,RC.5,RC.6}. Moreover, the zero- and first-order RC to the Casimir energy were developed in the system where the Lorentz symmetry was violated \cite{Reza.LV.}. The theory of relativity is the basis for the Quantum Field Theory\,(QFT) and implies the Lorentz symmetry to be fully conserved in QFT. However, some theories investigate models where the Lorentz symmetry is being violated. In the quantum gravity, Ho$\check{r}$ava--Lifshitz (HL) theory is a theory where breaks the Lorentz symmetry strongly. The HL gravity originated from a Lifshitz scalar field theory studied in the condensed matter physics as a description of tricritical phenomena involving spatially modulated phases \cite{Lifshitz.1,Lifshitz.2,hornreich}. Moreover, the Lifshitz-like theory was also studied in the framework of Maxwell's electromagnetic field\,\cite{horava} and scalar field theories \cite{Russo}. The Casimir energy for the Lifshitz scalar field was investigated previously \cite{Horava.Casimir.}, and its one-loop renormalization program was conducted in Ref. \cite{Russo}. In this work, we carried out the one-loop RC to the Casimir energy for a simple case of self-interacting Lifshitz-like scalar field, by mainly focusing on a special renormalization program. In the renormalization program followed by this study, the counterterm that usually eliminates the infinities originated from the bare parameters of the Lagrangian, is position-dependent. The merit of this type of counterterm in renormalizing the bare parameters of the Lagrangian compared to the earlier one made a controversy issue in recent studies \cite{1D.Reza}.
In the earlier works, the \emph{free counterterm} was used in any problem with any boundary condition imposed \cite{Graham.1,Graham.2,Graham.3,Graham.4,Graham.5,Graham.6}. Our meaning for the free counterterm is the one used for the Minkowski space. On the contrary, some studies argue when the non-trivial boundary condition or topology influences the quantum field, all elements of the renormalization program\,(\emph{e.g.}, the counterterms) should be consistent with it. Based on this argument, the use of one type of counterterm, without considering the boundary conditions imposed in the problem, may cause not all divergences owing to the bare parameters of the Lagrangian to be renormalized \cite{RC.2}. Hence, to resolve this possible problem, a systematic renormalization program was prescribed and conducted. The ensuing result of their renormalization program is finding the position-dependent counterterm consistent with the boundary condition imposed \cite{3D-Reza}. In this study, assuming the correctness of their hypothesis, we allowed the counterterms, to be automatically extracted from the renormalization program. It caused the resultant counterterms to be position-dependent, and they reflected the influence of the Lifshitz symmetry breaking. Using this type of counterterm, the vacuum energy of our system was systematically calculated up to the first-order of coupling constant. This renormalization program was successful, and its final solution was consistent with known physical principles.  Another important part of our calculation is the use of a method to remove the divergences appeared in the Casimir energy calculation. Since, in this paper, the RC of the Casimir energy for the Lifshitz-like scalar field with several types of boundary conditions (Dirichlet, Neumann, Mixed, and Periodic) in two cases (massive and massless scalar fields) was investigated, the appeared divergences were highly varied. Therefore, we needed to use a powerful regularization technique having the least amount of ambiguity or analytical continuation. Hence, the Box Subtraction Scheme (BSS) as the main regularization technique was used. This subtraction scheme is based on Boyer's method, and it can regularize the infinities without resorting to any analytic continuation \cite{boyer}. In this scheme, the vacuum energy of two different configurations with a similar nature is subtracted from each other. The parameters added by two configurations play the role of the regulator and a cutoff in some places of the calculation. Eventually, the sameness of configurations and their associated additional parameters cause the regularization of infinities in the calculation process to be conducted with more clarity.
\par
This paper follows the structure: In Section\,\ref{sec: model}, we introduce a model to execute the renormalization program in which the counterterm is position-dependent. In this model, how to deduce the counterterm and the vacuum energy was propounded. In Section\,\ref{sec:Radiative correction} and its entire subsections, the details of our calculation, including how to obtain the RC to the Casimir energy for the self-interacting massive/massless Lifshitz-like scalar field were presented. The paper finishes in Section\,\ref{sec:conclusion}, where the main results are summarized.
\section{The Model}\label{sec: model}
\begin{figure}[th]\centering\includegraphics[width=8cm]{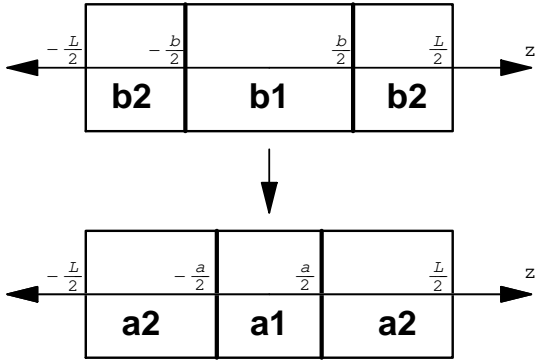}
\caption{\label{fig.1}
The geometry of the two different configurations whose energies are to be compared. The labels $\mathbf{a1}$, $\mathbf{b1}$, \textit{etc.}, denote the appropriate sections in each configuration separated by plates. The down configuration and the upper one are denoted by ``$\mathbf{A}$'' configuration and ``$\mathbf{B}$'' configuration, respectively\,\cite{RC.2}.}
\end{figure}

The Lagrangian of the Lifshitz scalar field takes the form of:
\begin{eqnarray}\label{lagrangian.form.1}
     \mathcal{L}=\frac{1}{2}&\Bigg[&(\partial_0\phi)^2-\ell^{2(\xi-1)}(\partial_{i}\phi)^{2\xi}-m_{0}^2\phi^2
     \nonumber\\&-&2\sum_{n=1}^{N_{\lambda}}\frac{\lambda_n}{(2n+2)!}\phi^{2n+2}
     \nonumber\\&-&\bigg(\alpha^2+2\sum_{n=1}^{N_\eta}\frac{\eta_n}{(2n)!}\phi^{2n}\bigg)(\partial_i\phi)^2\Bigg],
\end{eqnarray}
where $\ell$ is the parameter encoding the Lorentz symmetry breaking, and $\xi$ is the critical exponent. The form of Lagrangian shown in Eq. (\ref{lagrangian.form.1}) is the general form, and it was also introduced in previous studies \cite{Eune}. Indeed, it is not easy to calculate the RC to the Casimir energy for the scalar field defined in this form of Lagrangian. In this paper, we intend to consider a simpler case. Therefore, to neglect the last term in Eq.\,(\ref{lagrangian.form.1}), we set $\alpha=N_\eta=0$, and for more convenience, we also proposed $N_{\lambda}=1$. For any critical exponent $\xi$, the parameters $m_0$ and $\lambda_1$ are the bare mass of the field and bare coupling constant, respectively. To renormalize these bare parameters of the Lagrangian, after re-scaling the field $\phi=\sqrt{Z}\phi_r$, the Lagrangian form given in Eq. (\ref{lagrangian.form.1}) becomes:
\begin{eqnarray}\label{lagrangian.form.2}
      \mathcal{L}&=&\frac{1}{2}(\partial_0\phi_r)^2-\frac{1}{2}\ell^{2(\xi-1)}(\partial_{i}\phi_r)^{2\xi}
      -\frac{1}{2}m^2\phi_{r}^2\nonumber\\&&-\frac{\lambda}{4!}\phi_{r}^{4}+\frac{1}{2}\delta_{Z}(\partial_0\phi_r)^2
      -\frac{1}{2}\delta_{Z}\ell^{2(\xi-1)}(\partial_{i}\phi_r)^{2\xi} \nonumber\\&&-\frac{1}{2}\delta_{m}\phi_{r}^2-\frac{\delta_{\lambda}}{4!}\phi_{r}^{4},
\end{eqnarray}
where $Z=\delta_Z+1$ is called the field strength renormalization factor. Moreover, the parameters $\delta_m=Zm^2_0-m^2$ and $\delta_\lambda=Z^2\lambda_1-\lambda$ are the mass and coupling constant counterterms, respectively. For the above Lagrangian, the Feynman rules associated with the counterterms are written as follows:
\begin{eqnarray}\label{feynman.rule.counterterm}
   \raisebox{0mm}{\includegraphics[width=1cm]{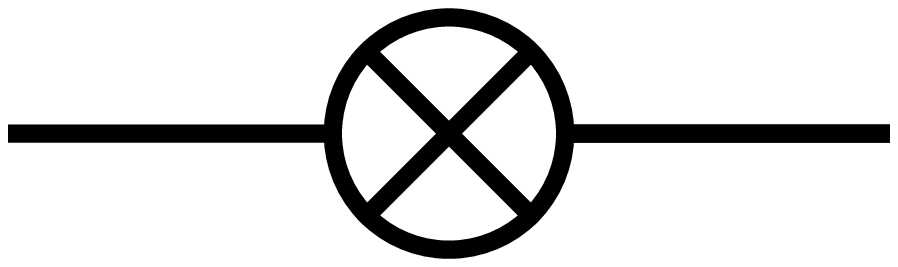}}&=&i\big[\ell^{2(\xi-1)}p^{2\xi}\delta_Z-\delta_m\big],\nonumber\\
   \raisebox{-2mm}{\includegraphics[width=0.7cm]{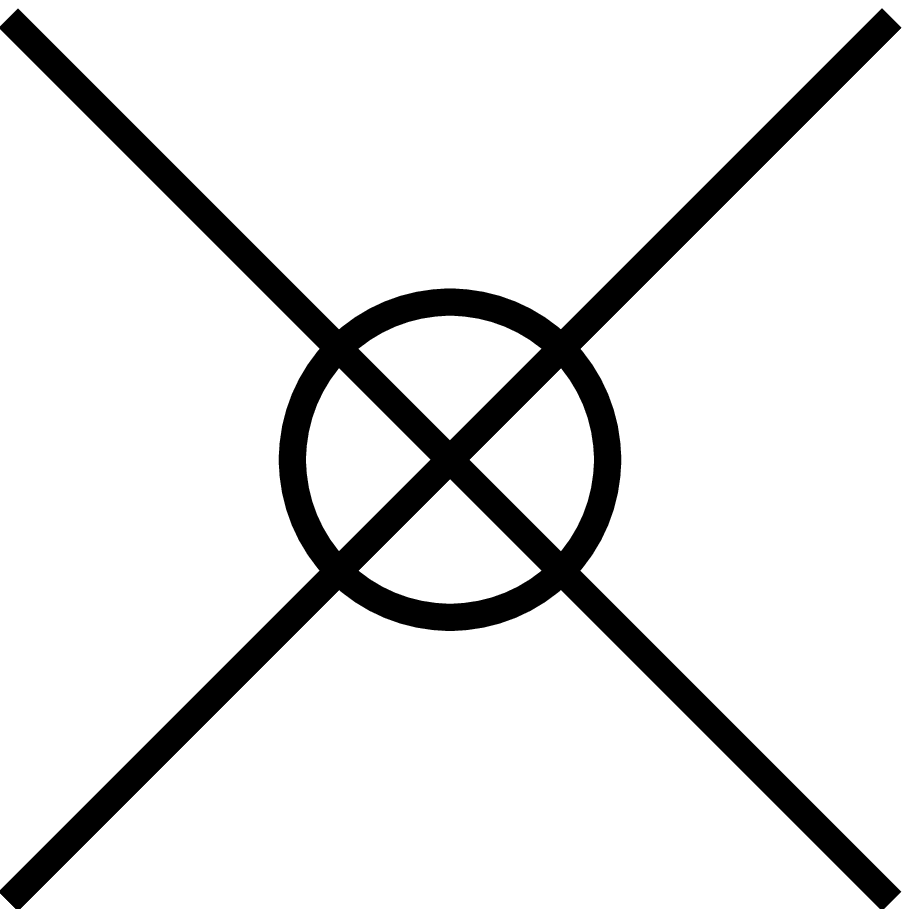}}&=&-i\delta_\lambda.
\end{eqnarray}
In the context of the renormalized perturbation theory, as indicated in Eq. (\ref{lagrangian.form.2}), we can symbolically represent the first few terms of the perturbation expansion of the two-point function by
\begin{equation}\label{twopoint.function.}
   \raisebox{-3mm}{\includegraphics[width=1.2cm]{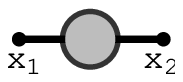}}=\raisebox{-1.5mm}{\includegraphics[width=1.2cm]{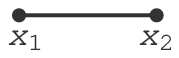}}
   +\raisebox{-2.5mm}{\includegraphics[width=1cm]{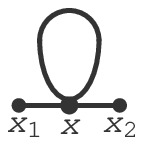}}+\raisebox{-2.5mm}{\includegraphics[width=1.2cm]{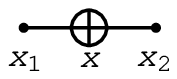}}\hspace{0cm},
\end{equation}
where $\raisebox{-2mm}{\includegraphics[width=1.2cm]{16.eps}}$ is the counterterm. To determine the values of the counterterm, the following form of renormalization conditions is needed:
\begin{eqnarray}\label{renormalization.conditions}
     \raisebox{-2mm}{\includegraphics[width=1cm]{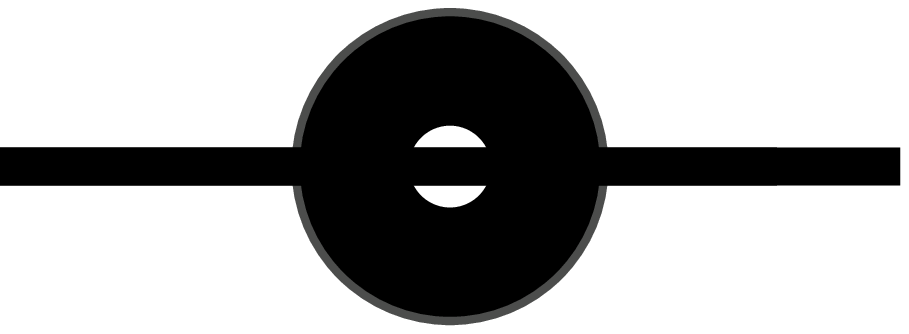}}&=&\frac{i}{\ell^{2(\xi-1)}p^{2\xi}-m^2}+\mbox{\tiny(the terms regular at $p^2=\mu^2$)},\nonumber\\
   \raisebox{-2mm}{\includegraphics[width=0.7cm]{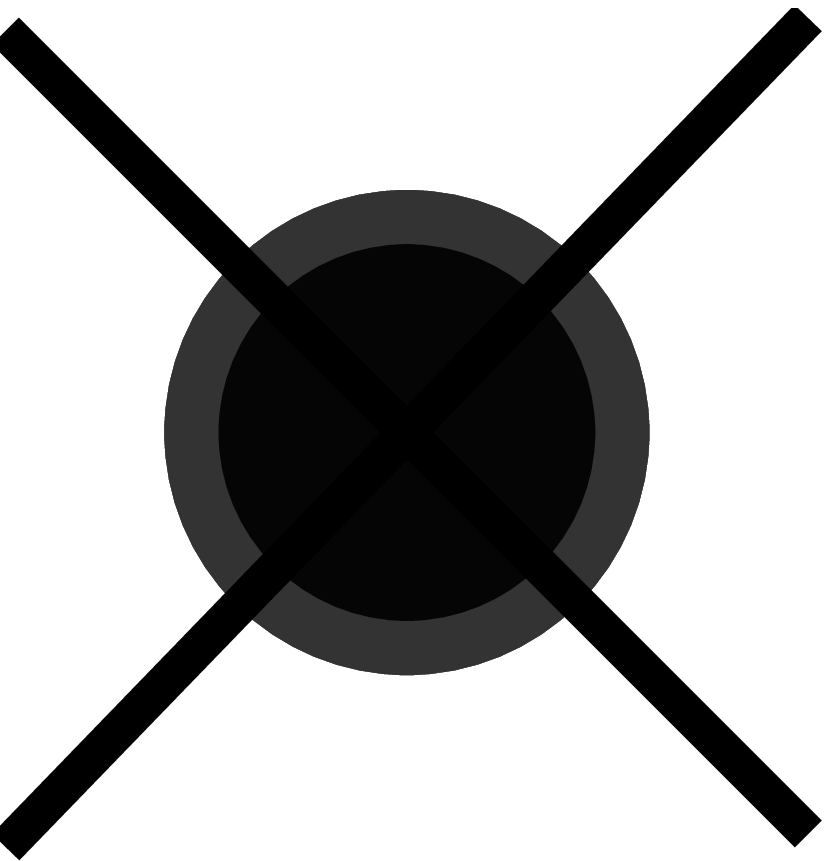}}&=&-i\lambda\hspace{2cm} \mbox{(at $s=4m^2$,$t=u=0$)}.
\end{eqnarray}
where $\mu=\sqrt[\xi]{m\ell}/\ell$. Moreover, the parameters $s$, $t$ and $u$ indicate the type of the channel. As known, the channel can be read from the form of the Feynman diagram, and each channel leads to the characteristic angular dependence of the cross section. Now, using Eq. (\ref{twopoint.function.}) and applying the renormalization condition given in Eq. (\ref{renormalization.conditions}) up to the first-order of coupling constant $\lambda$, the general expression for the mass counterterm becomes:
\begin{equation}\label{Counter-terms.}
     \delta_m(x)=\frac{-i}{2}\raisebox{-2mm}{\includegraphics[width=1cm]{15.eps}}=\frac{-\lambda}{2}G(x,x),
\end{equation}
where $G(x,x)$ is the Green's function. The Green's function used in this step, unlike some previous articles, is not the Green's function for the free space. Rather, this function is related to the space bounded by a boundary condition. This is exactly the point of difference between our work and the past works. In fact, the creation of the position-dependent countertem arises from the Green's function dependent on the boundary condition, causing the obtained counterterm to be compatible and reflecting the imposed boundary conditions of the problem. Up to the first-order of the coupling constant $\lambda$, the renormalization condition gives the zero value for counterterms $\delta_\lambda$ and $\delta_Z$. In the next step, we need to obtain an expression for the vacuum energy up to the first-order of coupling constant $\lambda$. Therefore, we have
\begin{eqnarray}\label{vacuum.energy.EXP1}
       E^{(1)}_{\mbox{\tiny vac.}}&=&i\int_{V} dV\bigg(\frac{1}{8} \raisebox{-7mm}{\includegraphics[width=0.5cm]{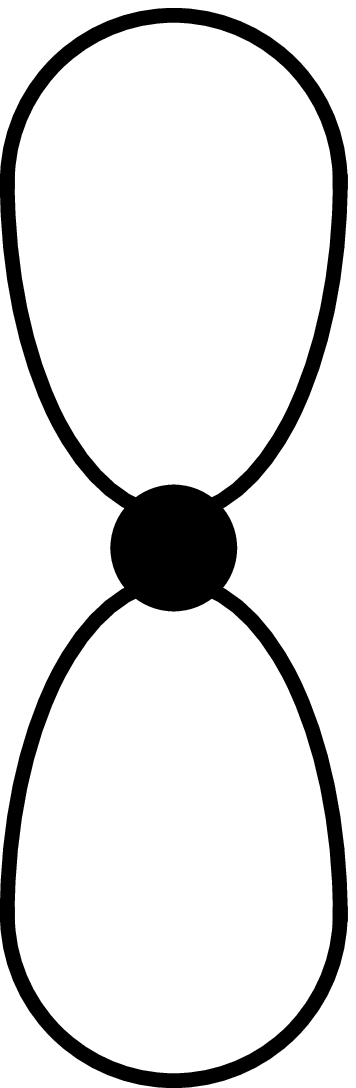}}
       +\frac{1}{2}\raisebox{-1mm}{\includegraphics[width=0.5cm]{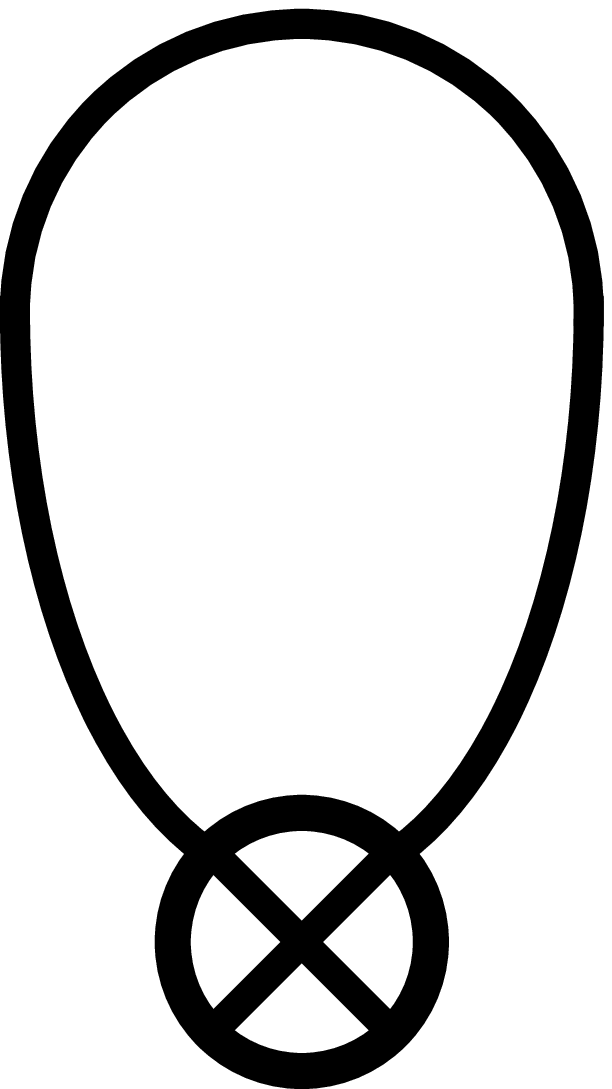}}+...\bigg)
       \\&=&i \int_{V}dV\bigg(\frac{-i\lambda}{8}G^2(x,x)-\frac{-i}{2}\delta_m(x)G(x,x)\bigg),\nonumber
\end{eqnarray}
where $V$ denotes a volume that is made by two large parallel plates with a specific distance. The superscript $(1)$ on the vacuum energy indicates the first-order of the coupling constant $\lambda$. Combination of Eqs. (\ref{Counter-terms.}) and (\ref{vacuum.energy.EXP1}) for the region $\mathbf{a1}$ of Fig. (\ref{fig.1}) leads to:
\begin{eqnarray}\label{vacuum.energy.EXP2}
  E^{(1)}_{\mbox{\tiny vac.}}(a)=\frac{-\lambda}{8}\int_{V}dV G^2(a;x,x).
\end{eqnarray}
where $G(a;x,x)$ is the Green's function related to the region $\mathbf{a1}$ of Fig. (\ref{fig.1}). By having the Green's function related to each region of Fig. (\ref{fig.1}), the vacuum energy of each region can be obtained separately. In the earlier studies, to obtain the Casimir energy, the vacuum energy of the system in the presence and absence of non-trivial boundary conditions were subtracted with each other. In this paper, however, we followed a different manner. To obtain the first-order RC to the Casimir energy, we defined two similar configurations as what displayed in Fig. (\ref{fig.1}). In this figure, two infinite parallel plates\,(with distance $a$) were placed within two other plates\,(with distance $L>a$). In fact, two outer plates with the distance $L$ played the role of box for two inner ones. We then constructed a similar configuration of plates with distances $b>a$. Next, we subtracted the vacuum energies of these two configurations. Finally, to obtain the Casimir energy for the original configuration\,(two large parallel plates with the distance $a$), we let $L$ and then $b$ go to infinity. This formalism to obtain the Casimir energy, which is based on Boyer's method is categorised as a regularization technique\,\cite{boyer}. Heretofore, this technique of regularization to extract of the Casimir energy was called the Box Subtraction Scheme (BSS)\,\cite{RC.2}. We beilieved that the vacuum energy of the Minkowski space was simulated by the configuration ``$\mathbf{B}$'' displayed in Fig. (\ref{fig.1}), when its sizes ($L$ and $b$) tended to infinity. The goodness of this simulation is the addition of more regulators in the subtraction process of the vacuum energy. It could provide a situation in which all infinities are eliminated with the least number of possible ambiguities. In BSS, as a regularization technique, the Casimir energy is usually defined as:
\begin{eqnarray}\label{BSS.Definition}
    E_{\mbox{\tiny Cas.}}=\lim_{b\to\infty}\Big[\lim_{L\to\infty}
    \big[E_{\mathbf{A}}-E_{\mathbf{B}}\big]\Big],
\end{eqnarray}
where $E_\mathbf{A}=E_{\mathbf{a1}}+2E_{\mathbf{a2}}$ and $E_\mathbf{B}=E_{\mathbf{b1}}+2E_{\mathbf{b2}}$ are the vacuum energies of configuration $\mathbf{A}$ and $\mathbf{B}$, respectively. According to Eqs. (\ref{vacuum.energy.EXP2}) and (\ref{BSS.Definition}), to initiate the Casimir energy calculation, we need first to obtain the Green's function.  To achieve this quantity, the computations were followed in the next subsection.
\subsection{Green's Function}\label{subsec: Green's Function}
In $3+1$ dimensions, the modified Klein-Gordon equation associated with the non-interacting Lifshitz scalar field ($\lambda_1=0$) reads:
\begin{eqnarray}\label{Klein-Gordon.Eq}
      \Big[\partial_0^2+\ell^{2(\xi-1)}(-1)^{\xi}(\partial^2_x+\partial^2_y+\partial^2_z)^{\xi}+m^2\Big]\phi(x)=0.\nonumber\\
\end{eqnarray}
For two large parallel plates separated by an orthogonal $z$ axis with a small distance $a$\,(region $\mathbf{a1}$ displayed in Fig. (\ref{fig.1})), the Dirichlet Boundary Condition (DBC) reads:
\begin{eqnarray}\label{Dirichlet.BC.}
      \phi\Big|_{z=\frac{\pm a}{2}}=0.
\end{eqnarray}
Applying the above boundary condition on the solution of the equation of motion, given in Eq. (\ref{Klein-Gordon.Eq}), yields the eigenfunction expression. Afetrwards, adopting the standard procedure in obtaining the Green's function expression for the region $\mathbf{a1}$ together with applying the Wick rotation leads to (for more details see please Appendix \ref{Appendix.Green.Function}):
\begin{eqnarray}\label{Green's.Func.Dirichlet}
G_{\mathcal{D}}(a;x,x')&=&\frac{2}{a}\int\frac{d^3k}{(2\pi)^3}\nonumber\\&&\times\sum_{n=1}^{\infty}\frac{\begin{array}{c}
                                                                      e^{-\omega(t-t')}e^{i\mathbf{k}\cdot(\mathbf{X}-\mathbf{X}')}\\
                                                                      \times\sin[\frac{n\pi}{a}(z+\frac{a}{2})]\sin[\frac{n\pi}{a}(z'+\frac{a}{2})]
                                                                    \end{array}}{\omega^2+\omega^{\mbox{\tiny$(\mathcal{D})$}}_{\xi,n}(\mathbf{k})^2},\nonumber\\
\end{eqnarray}
where the wave-vector $\mathbf{k}=(k_x,k_y)$ and $k=(\omega,\mathbf{k})$. Additionally, the coordinate $\mathbf{X}=(x,y)$ and the subscript $\mathcal{D}$ denotes DBC. According to the DBC, the allowed wave numbers read:
\begin{eqnarray}\label{wave.number.Dirichlet}
       \omega^{\mbox{\tiny$(\mathcal{D})$}}_{\xi,n}(\mathbf{k})
       =\bigg[\ell^{2(\xi-1)}\bigg(\mathbf{k}^2+\Big(\frac{n\pi}{a}\Big)^2\bigg)^{\xi}+m^2\bigg]^{1/2}.
\end{eqnarray}
To obtain the Green's function expression related to the case of the problem in which the Lifshitz-like scalar field is bounded by the Mixed Boundary Condition (MBC) between two parallel plates, we reiterate the scenario considered in the Appendix \ref{Appendix.Green.Function}. In applying MBC, the Dirichlet and Neumann boundary conditions are used simultaneously. Precisely, to apply this type of boundary condition on two parallel plates, the DBC should be satisfied on the left plate\,(\emph{e.g.} the plate placed on $z=-a/2$) and on the opposite side, the plate placed on the right ($z=a/2$) satisfies the Neumann Boundary Condition (NBC). It should be noted that for the reverse order in applying the Dirichlet and Neumann boundary conditions on the plates, the vacuum energy expression remains unchanged. For the original region $\mathbf{a1}$ displayed in Fig. (\ref{fig.1}), after performing the standard computations in determining the Green's function expression, this expression for the massive Lifshitz - like scalar field confined with MBC between two plates located on $z=\pm a/2$ in $3+1$ dimensions after the Wick rotation becomes:
\begin{eqnarray}\label{Green's.Func.Mixed}
     G_{\mathcal{M}}(a;x,x')&=&\frac{1}{a}\int\frac{d^3k}{(2\pi)^3}\\&&\hspace{0.2cm}\times\sum_{n=0}^{\infty}\frac{\begin{array}{c}
                                                                      e^{-\omega(t-t')}e^{i\mathbf{k}\cdot(\mathbf{X}-\mathbf{X}')}
                                                                      \\ \times[\sin(k_{n}z)+(-1)^{n}\cos(k_{n}z)]\\
                                                                      \times[\sin(k_{n}z')+(-1)^{n}\cos(k_{n}z')]
                                                                    \end{array}}{\omega^2+\omega^{\mbox{\tiny$(\mathcal{M})$}}_{\xi,n}(\mathbf{k})^2},\nonumber
\end{eqnarray}
where $\omega^{\mbox{\tiny$(\mathcal{M})$}}_{\xi,n}(\mathbf{k})=[\ell^{2(\xi-1)}(\mathbf{k}^2+k_n^2)^{\xi}+m^2\big]^{1/2}$ is the allowed wave number and $k_n=\frac{(2n+1)\pi}{2a}$. Now, with the Green's functions obtained in Eqs. (\ref{Green's.Func.Dirichlet}) and (\ref{Green's.Func.Mixed}), and using Eq. (\ref{vacuum.energy.EXP2}), the vacuum energy expression for the problem in the case of DBC/MBC is achievable. We followed this subject in the next section.
\section{Radiative Correction}\label{sec:Radiative correction}
In this section, we presented the computation of RC to the Casimir energy for the massive/massless Lifshitz-like scalar field confined with DBC/MBC between two parallel plates in $3+1$ dimensions. This calculation was divided into the following subsections according to the type of boundary conditions (Dirichlet and Mixed).
\subsection{Dirichlet Boundary Condition}\label{subsec: DBC.RC}
According to the definition form of the Casimir energy presented in Eq. (\ref{BSS.Definition}), the vacuum energy of all regions of Fig. (\ref{fig.1}) is required. Hence, we put the distance of the plates corresponding to each created region shown in Fig. (\ref{fig.1}), into Eq. (\ref{Green's.Func.Dirichlet}). This substitution provides us with the Green's function expression of each region effortlessly. By substituting the Green's function expression related to each region of Fig. (\ref{fig.1}) in Eq. (\ref{vacuum.energy.EXP2}), the vacuum energy of each region is achievable. Now, using the Casimir energy definitions introduced in Eq. (\ref{BSS.Definition}), the subtraction of the vacuum energy density of two configurations displayed in Fig. (\ref{fig.1}) is obtained as follows:
\begin{eqnarray}\label{Delta.vac.DBC.1}
     \Delta \mathcal{E}_{\mbox{\tiny$\mathcal{D}$, vac}}^{(1)[m]}&=&\frac{-\lambda \pi^2\ell^{2(1-\xi)}}{8(2\pi)^6a}
       \Bigg\{\sum_{n,n'=1}^{\infty}f^{\mbox{\tiny[$m$]}}_{n}(a,\xi)f^{\mbox{\tiny[$m$]}}_{n'}(a,\xi)\nonumber\\
     &&+\frac{1}{2}\sum_{n=1}^{\infty}\Big[f^{\mbox{\tiny[$m$]}}_{n}(a,\xi)\Big]^2\Bigg\}
      +2\times\{a\to\frac{L-a}{2}\}\nonumber\\
     &&-\{a\to b\}-2\times\{a\to\frac{L-b}{2}\},
\end{eqnarray}
where the function $f^{\mbox{\tiny[$m$]}}_n(a,\xi)$ is
\begin{eqnarray}\label{f(a,n).DBC.RC}
     f^{\mbox{\tiny[$m$]}}_{n}(a,\xi)=\int\int\frac{dk_xdk_y}{((k^2_x+k^2_y+(\frac{n\pi}{a})^2)^\xi+\mu^{2\xi})^{1/2}}.
\end{eqnarray}
The superscript $[m]$ denotes the massive case of the Lifshitz-like scalar field. To reduce the complexity of the summation forms of Eq. (\ref{Delta.vac.DBC.1}) and to convert them into the integration form, the following equation named as Abel-Plana Summation Formula (APSF) was used \cite{A.Saharian},
\begin{eqnarray}\label{APSF.INTEGER}
      \sum_{n=1}^{\infty}\mathcal{F}(n)&=&\frac{-1}{2}\mathcal{F}(0)+\int_{0}^{\infty}\mathcal{F}(x)dx\nonumber\\
      &&\hspace{1.2cm}+i\int_{0}^{\infty}\frac{\mathcal{F}(it)-\mathcal{F}(-it)}{e^{2\pi t}-1}dt.
\end{eqnarray}
The second term on the right hand side of Eq. (\ref{APSF.INTEGER}) is called the \emph{integral term}, and the third one is usually called the \emph{branch-cut term}. Applying the above form of APSF on summations written in Eq. (\ref{Delta.vac.DBC.1}), we obtain
\begin{eqnarray}\label{Delta.vac.DBC.2}
      \Delta \mathcal{E}_{\mbox{\tiny$\mathcal{D}$, vac}}^{(1)[m]}&=&\frac{-\lambda \pi^2\ell^{2(1-\xi)}}{8(2\pi)^6a}
      \Bigg\{\Big[\frac{-1}{2}f^{\mbox{\tiny[$m$]}}_{0}(a,\xi)\nonumber\\
     &&\hspace{-0cm}+\underbrace{\int_{0}^{\infty}f^{\mbox{\tiny[$m$]}}_x(a,\xi)dx}_{\mathcal{I}_1(a;\infty)}
     +B_1(m,a;\xi)\Big]^2
     \nonumber\\&&\hspace{-0cm}-\frac{1}{4}f^{\mbox{\tiny[$m$]}}_0(a,\xi)^2
     +\underbrace{\frac{1}{2}\int_{0}^{\infty}f^{\mbox{\tiny[$m$]}}_{x}(a,\xi)^2dx}_{\mathcal{I}_2(a;\infty)}
     \nonumber\\
     &&\hspace{-0cm}+\frac{1}{2}B_2(m,a;\xi)\Bigg\}+2\times\{a\to\frac{L-a}{2}\}\nonumber\\
     &&\hspace{-0cm}-\{a\to b\}-2\times\{a\to\frac{L-b}{2}\},
\end{eqnarray}
where $B_1(m,a;\xi)$ and $B_2(m,a;\xi)$ are the branch-cut terms of APSF, and all details for their computations are presented in Appendix \ref{Appendix.1}. For $\xi\leq3$, the integral term $\mathcal{I}_1(a;\infty)$ is explicitly divergent. Therefore, any term multiplied by it will also be divergent, and needs to be regularized, and ultimately their emerged infinities must be removed. For this purpose, the extended form of the first bracket in Eq. (\ref{Delta.vac.DBC.2}) was obtained. So, we have:
\begin{eqnarray}\label{Delta.vac.DBC.2..5}
\Delta \mathcal{E}_{\mbox{\tiny$\mathcal{D}$, vac}}^{(1)[m]}&=&\frac{-\lambda \pi^2\ell^{2(1-\xi)}}{8(2\pi)^6a}\nonumber\\&&\hspace{-0.4cm}\times
      \Bigg\{\mathcal{I}_1(a;\infty)^2-\mathcal{I}_1(a;\infty)f^{\mbox{\tiny[$m$]}}_{0}(a,\xi)\nonumber\\
     &&\hspace{-0.4cm}
     +2\mathcal{I}_1(a;\infty)B_1(m,a;\xi)-f^{\mbox{\tiny[$m$]}}_{0}(a,\xi)B_1(m,a;\xi)
      \nonumber\\&&\hspace{-0.4cm}+B_1(m,a;\xi)^2
     +\mathcal{I}_2(a;\infty)
     \nonumber\\
     &&\hspace{-0.4cm}+\frac{1}{2}B_2(m,a;\xi)\Bigg\}+2\times\{a\to\frac{L-a}{2}\}\nonumber\\
     &&\hspace{-0.4cm}-\{a\to b\}-2\times\{a\to\frac{L-b}{2}\},
\end{eqnarray}
In the integral $\mathcal{I}_1(a;\infty)$, we set $\mathcal{Z}=\frac{x\pi}{a}$. This change of variables leads to:
\begin{eqnarray}\label{I1.more discussion}
       \mathcal{I}_1(a;\infty)=\frac{a}{2\pi}\int\int\int
       \frac{dk_xdk_yd\mathcal{Z}}{((k^2_x+k^2_y+\mathcal{Z}^2)^\xi+\mu^{2\xi})^{1/2}}.
\end{eqnarray}
This change of variables help to regulate infinities originating from the integral terms, so that they would be removed via the subtraction procedure defined by BSS. For instance, the elimination process of infinities for the term $\mathcal{I}^2_1$, via the use of BSS, can be written as:
\begin{eqnarray}\label{I.Disappear.1}
       &&\frac{1}{a}\mathcal{I}^2_1(a;\infty)+\frac{4}{L-a}\mathcal{I}^2_1(\frac{L-a}{2};\infty)-\{a\to b\}\nonumber\\
       &&\hspace{2cm}=\Big[a+2\frac{L-a}{2}-b-2\frac{L-b}{2}\Big]\nonumber\\
       &&\hspace{2.5cm}\times\Bigg(\frac{1}{2\pi}\int_{0}^{\infty}\frac{4\pi\rho^2d\rho}{(\rho^{2\xi}
       +\mu^{2\xi})^{1/2}}\Bigg)^2=0,\nonumber\\
\end{eqnarray}
where the change of variables $\rho^2=k_x^2+k_y^2+\mathcal{Z}^2$ was used. Analogously, the elimination process for the integral term $\mathcal{I}_2(a;\infty)$ occured as follows:
\begin{eqnarray}\label{I.Disappear.2}
      \frac{1}{a}\mathcal{I}_2(a;\infty)&+&\frac{4}{L-a}\mathcal{I}_2(\frac{L-a}{2};\infty)-\{a\to b\}\nonumber\\&&\hspace{-2cm}
      =\Big[\frac{1}{a}\frac{a}{\pi}+\frac{4}{L-a}\frac{L-a}{2\pi}-\{a\to b\}\Big]\nonumber\\&&\hspace{-1.5cm}
      \times\int_{0}^{\infty}d\mathcal{Z}\Bigg(\int_{0}^{\infty}
      \frac{2\pi\mathcal{\rho'}d\mathcal{\rho'}}{((\mathcal{\rho'}^2+\mathcal{Z}^2)^{\xi}
      +\mu^{2\xi})^{1/2}}\Bigg)^2=0,\nonumber\\
\end{eqnarray}
where ${\rho'}^2=k_x^2+k_y^2$ and $\mathcal{Z}=\frac{x\pi}{a}$. For the cross-term $\frac{1}{a}f^{\mbox{\tiny[m]}}_{0}(a,\xi)\mathcal{I}_1(a;\infty)$ written in Eq. (\ref{Delta.vac.DBC.2..5}), we have:
\begin{eqnarray}\label{cross.term.f_0DAR_I}
           &&\frac{1}{a}f^{\mbox{\tiny[$m$]}}_{0}(a,\xi)\mathcal{I}_1(a;\infty)
           +\frac{4}{L-a}f^{\mbox{\tiny[$m$]}}_{0}(\frac{L-a}{2},\xi)\mathcal{I}_1(\frac{L-a}{2};\infty)
           \nonumber\\&&-\{a\to b\}
           =\Big[\frac{1}{a}\frac{a}{\pi}+\frac{4}{L-a}\frac{(L-a)}{2\pi}-\{a\to b\}\Big]\nonumber\\&&\hspace{0.5cm}
           \times\int_{0}^{\infty}
      \frac{2\pi\mathcal{\rho}d\mathcal{\rho}}{(\mathcal{\rho}^{2\xi}
      +\mu^{2\xi})^{1/2}}\Bigg(\int_{0}^{\infty}
           \frac{2\pi\mathcal{\rho'}^2d\mathcal{\rho'}}{(\mathcal{\rho'}^{2\xi}+\mu^{2\xi})^{1/2}}\Bigg)=0\nonumber\\
\end{eqnarray}
Based on the above three equations, Eq. (\ref{Delta.vac.DBC.2..5}) is converted to:
\begin{eqnarray}\label{Delta.vac.DBC.3}
       \Delta \mathcal{E}_{\mbox{\tiny$\mathcal{D}$, vac}}^{(1)[m]}&=&\frac{-\lambda \pi^2\ell^{2(1-\xi)}}{8(2\pi)^6a}\Big[-f^{\mbox{\tiny[$m$]}}_{0}(a,\xi)B_1(m,a;\xi)\nonumber\\
       &+&2\mathcal{I}_1(a;\infty)B_1(m,a;\xi)+B_1(m,a;\xi)^2\nonumber\\
       &+&\frac{1}{2}B_2(m,a;\xi)\Big]+2\times\{a\to\frac{L-a}{2}\}
       \nonumber\\&-&\{a\to b\}-2\times\{a\to\frac{L-b}{2}\}.
\end{eqnarray}
For all values of $\xi>2$, the function $f^{\mbox{\tiny[m]}}_{0}(a,\xi)$ is convergent, and its value is:
\begin{eqnarray}\label{f_value}
       f^{\mbox{\tiny[$m$]}}_{0}=\frac{\sqrt{\pi}\Gamma\left(\frac{\xi-2}{2\xi}\right)\Gamma\left(1+\frac{1}{\xi}\right)}
{\mu^{\xi-2}}.
\end{eqnarray}
However, for $\xi=1$ and $\xi=2$, this function does not converge, causing the first term on the right-hand side of Eq. (\ref{Delta.vac.DBC.3}) to be divergent. To remove its infinities, in addition to BSS as our main regularization scheme, it is necessary to employ supplementary regularization like the cutoff regularization technique. Therefore, we started with $\xi=1$ and we replaced the upper limit of the integral $f^{\mbox{\tiny[m]}}_{0}(a,1)$ with a cutoff. Therefore, by calculating the integral, we obtained:
\begin{eqnarray}\label{f(a,0).DBC.cutoff.1}
        f^{\mbox{\tiny[$m$]}}_0(a,1)=\int_{0}^{\Lambda_\mathbf{a1}}\frac{2\pi\rho'd\rho'}{\sqrt{{\rho'}^{2}+\mu^{2}}}
        =2\pi(\sqrt{\Lambda_\mathbf{a1}^2+\mu^2}-\mu).\nonumber\\
\end{eqnarray}
We then expanded the result of integration at the infinite limit of cutoff $\Lambda_\mathbf{a1}$. This expansion caused the infinite part of the integration to be manifested. This scenario should be conducted for similar terms of Eq. (\ref{Delta.vac.DBC.3}) by various cutoffs. Ultimately, for the first term of Eq. (\ref{Delta.vac.DBC.3}), it can be written:
\begin{eqnarray}\label{removing.f DAR B_1}
    &&\hspace{-0.6cm}\Big[\frac{-f^{\mbox{\tiny[$m$]}}_0(a,1)}{a}B_1(m,a;1)+2\times\{a\to\frac{L-a}{2}\}\Big]-\{a\to b\}\nonumber\\&=&-2\pi\mu[\frac{B_1(m,a;1)}{a}+\frac{4B_1(m,\frac{L-a}{2};1)}{L-a}]-\{a\to b\}\nonumber\\&+&
    2\pi[\frac{B_1(m,a;1)}{a}\Lambda_\mathbf{a1}+\frac{4B_1(m,\frac{L-a}{2};1)}{L-a}\Lambda_\mathbf{a2}]-\{a\to b\}\nonumber\\&+&\mathcal{O}(1/\Lambda).
\end{eqnarray}
All terms in the second bracket on the right-hand side of Eq. (\ref{removing.f DAR B_1}) are divergent due to the cutoff value. Adjustment of the cutoffs as the following form will remove all divergent parts from the above expansion,
\begin{eqnarray}\label{adjusting.cutoffs}
       \frac{\Lambda_\mathbf{a1}}{\Lambda_\mathbf{b1}}&=&\frac{aB_1(m,b;1)}{bB_1(m,a;1)},\nonumber\\
       \frac{\Lambda_\mathbf{a2}}{\Lambda_\mathbf{b2}}&=&
       \frac{(L-a)B_1(m,\frac{L-b}{2};1)}{(L-b)B_1(m,\frac{L-a}{2};1)}
\end{eqnarray}
As a result, the only remained finite contribution from the function $f^{\mbox{\tiny[m]}}_0(a,1)$ is equal to $-2\pi\mu$. In the case of $\xi=2$, the reiteration of the computing scenario conducted in Eqs. (\ref{f(a,0).DBC.cutoff.1}) to (\ref{adjusting.cutoffs}) leads to the finite contribution for the function $f^{\mbox{\tiny[m]}}_0(a,2)$ as $\pi\ln2$. In fact, for any values of $\xi$, no divergent contribution remains from the function $f^{\mbox{\tiny[m]}}_0$ in Eq. (\ref{Delta.vac.DBC.3}). Only, the remained finite parts from the function $f^{\mbox{\tiny[m]}}_{0}(a,\xi)$ associated with each value of $\xi$ are:
\begin{eqnarray}\label{f(a,0).DBC.cutoff.2}
&&f^{\mbox{\tiny[$m$]}}_{0}(a,\xi)\buildrel \Lambda_{a}\to\infty\over\longrightarrow\nonumber\\
                           &&\hspace{1cm}\tilde{F}^{\mbox{\tiny$[m]$}}(\xi)=\left\{
                          \begin{array}{ll}
                           -2\pi\mu, & \hspace{0.7cm}\hbox{\small$\xi=1$;} \\
                           \pi\ln2, & \hspace{0.7cm}\hbox{\small$\xi=2$;}\\
                           \frac{\sqrt{\pi}\Gamma\left(\frac{\xi-2}{2\xi}\right)\Gamma\left(1+\frac{1}{\xi}\right)}
                           {\mu^{\xi-2}}& \hspace{0.7cm}\hbox{\small other $\xi$.}
                          \end{array}\right. \nonumber\\
\end{eqnarray}
For all $\xi>3$, the integral $\mathcal{I}_1$ is convergent. Its value is:
\begin{eqnarray}\label{I_1.Value.for.xi>3}
       \mathcal{I}_1(a;\infty)=\frac{2a\Gamma\left(\frac{\xi-3}{2\xi}\right)
       \Gamma\left(1+\frac{3}{2\xi}\right)}{3\sqrt{\pi}\mu^{\xi-3 }}.
\end{eqnarray}
However, for $\xi=1,2$ and $3$, this function tends to infinity. It makes that the second term on the right-hand side of Eq. (\ref{Delta.vac.DBC.3}) is to be divergent. Like the first term on the right-hand side of Eq. (\ref{Delta.vac.DBC.3}), we should get rid of this divergent contribuation. Therefore, we started with $\xi=1$, and replaced the upper limit of integral $\mathcal{I}_1$ with a cutoff. The cutoff ${\Lambda'}_\mathbf{a1}$ was selected for the upper limit of $\mathcal{I}_1$ in the term related to regions $\mathbf{a1}$ of Fig. (\ref{fig.1}). Similarly, the cutoffs ${\Lambda'}_\mathbf{a2}$, ${\Lambda'}_\mathbf{b1}$, and ${\Lambda'}_\mathbf{b2}$ should be replaced on the upper limit of the integral $\mathcal{I}_1$ related to the other regions in Eq. (\ref{Delta.vac.DBC.3}). Then, by calculating the integral up to the cutoff value, the integration result was expanded at the infinite limit of the cutoff. Only for the integral $\mathcal{I}_1(a;{\Lambda'_{\mathbf{a1}}})$, we obtained:
\begin{eqnarray}\label{removing.I}
\hspace{-0.6cm}&&\mathcal{I}_1(a;{\Lambda'_{\mathbf{a1}}})=\frac{a}{2\pi}\int_{0}^{{\Lambda'_{\mathbf{a1}}}}
\frac{4\pi\rho^2d\rho}{\sqrt{\rho^{2\xi}+\mu^{2\xi}}}\buildrel{{\Lambda'_{\mathbf{a1}}\to\infty}}\over\longrightarrow\nonumber\\
&&a\mu^2(\frac{1}{2}-\ln2)+a{{\Lambda'}^2_{\mathbf{a1}}}+a\mu^2\ln\Big(\frac{{\Lambda'_{\mathbf{a1}}}}{\mu}\Big)+\mathcal{O}\Big(\frac{1}{{\Lambda'_{\mathbf{a1}}}}\Big).\nonumber\\
\end{eqnarray}
We also performed the same process for all similar terms written in Eq. (\ref{Delta.vac.DBC.3}). Now, by substituting the expanded form of the integral $\mathcal{I}_1$ in the second term of Eq. (\ref{Delta.vac.DBC.3}), and by adjusting the cutoffs as the following form, all divergences originating from the integral $\mathcal{I}_1$ will be eliminated:
\begin{eqnarray}\label{adjusting.cutoff.for.I}
\frac{\frac{{{\Lambda'}^2_\mathbf{a1}}}{\mu^2}+\ln({\Lambda'_\mathbf{a1}}/\mu)}{\frac{{{\Lambda'}^2_{\mathbf{b1}}}}{\mu^2}+\ln({\Lambda'_\mathbf{b1}}/\mu)}
&=&\frac{bB_1(m,b;1)}{aB_1(m,a;1)}\nonumber\\
\frac{\frac{{{\Lambda'}^2_{\mathbf{a2}}}}{\mu^2}+\ln({{\Lambda'}_\mathbf{a2}}/\mu)}{\frac{{{\Lambda'}^2_{\mathbf{b2}}}}{\mu^2}+\ln({\Lambda'_\mathbf{b2}}/\mu)}
&=&\frac{(L-b)B_1(m,\frac{L-b}{2};1)}{(L-a)B_1(m,\frac{L-a}{2};1)}.
\end{eqnarray}
Therefore, the only finite contribution remained in Eq. (\ref{Delta.vac.DBC.3}) for the integral $\mathcal{I}_1(a;\infty)$ is $a\mu^2(\frac{1}{2}-\ln2)$. As with the calculation procedure conducted for $\xi=1$, calculations can also be performed for $\xi=2$ and $3$. In each case, the final remaining finite contribution is defferent. Consequently, we may declare the finite contribution of the integral $\mathcal{I}_1$ for each value of $\xi$ as follows:
\begin{eqnarray}\label{Finite.remained.I_1(a)}
        &&\mathcal{I}_1(a;\Lambda'_\mathbf{a1})\buildrel\Lambda'_\mathbf{a1}\to\infty\over{\longrightarrow}\nonumber\\
        &&\hspace{1.3cm}\tilde{\mathcal{I}}(a)=\left\{
                          \begin{array}{ll}
                           a\mu^2(\frac{1}{2}-\ln2), & \hspace{0.6cm}\hbox{\small$\xi=1$;} \\
                             \frac{2a\mu}{3\sqrt{\pi}}\Gamma(-1/4)\Gamma(7/4), & \hspace{0.6cm}\hbox{\small$\xi=2$;}\\
                             \frac{2a}{3}\ln2, & \hspace{0.6cm}\hbox{\small$\xi=3$;}\\
                             \frac{2a\Gamma\left(\frac{\xi-3}{2\xi}\right)
                             \Gamma\left(1+\frac{3}{2\xi}\right)}{3\sqrt{\pi}\mu^{\xi-3 }},&\hspace{0.6cm}\mbox{\small
                            $\xi>3$.}
                          \end{array}\right. \nonumber\\
\end{eqnarray}
Ultimately, using Eqs. (\ref{Delta.vac.DBC.3}), (\ref{f(a,0).DBC.cutoff.2}), (\ref{Finite.remained.I_1(a)}) and (\ref{Calculation.B_2(a).No1}), we can express only finite parts remained from Eq. (\ref{Delta.vac.DBC.3}) as follows:
\begin{eqnarray}\label{Delta.vac.DBC.4}
       \Delta \mathcal{E}_{\mbox{\tiny$\mathcal{D}$, vac}}^{(1)[m]}&=&\frac{-\lambda \pi^2\ell^{2(1-\xi)}}{8(2\pi)^6a}B_1(m,a;\xi)      \nonumber\\&\times&\Big[B_1(m,a;\xi)+2\tilde{\mathcal{I}}(a)-\tilde{F}^{\mbox{\tiny$[m]$}}(\xi)\Big]
\nonumber\\&+&2\times\{a\to\frac{L-a}{2}\}-\{a\to b\}\nonumber\\&-&2\times\{a\to\frac{L-b}{2}\},
\end{eqnarray}
In the final step, the limits $b,L\to\infty$ according to Eq. (\ref{BSS.Definition}) should be applied. This limit leads the values of all branch-cut terms related to regions $\mathbf{a2}$, $\mathbf{b1}$ and $\mathbf{b2}$ to be vanished. Therefore, the final expression for RC to the Casimir energy density of the massive self-interacting Lifshitz-like scalar field confined with DBC between two parallel plates with distance $a$ in $3+1$ dimensions becomes:
\begin{eqnarray}\label{DBC.Casimir.En}
       \mathcal{E}_{\mbox{\tiny $\mathcal{D},\xi$}}^{\mbox{\tiny$(1)$Cas.}}&=&\frac{-\lambda \pi^2\ell^{2(1-\xi)}}{8(2\pi)^6a}      B_1(m,a;\xi)\Big[B_1(m,a;\xi)\nonumber\\&&\hspace{0.5cm}+2\tilde{\mathcal{I}}(a)-\tilde{F}^{\mbox{\tiny$[m]$}}(\xi)\Big].
\end{eqnarray}
We recall that the calculation of the branch-cut term $B_1(m,a;\xi)$ was expressed in Appendix \ref{Appendix.1}. One of the momentous limits that is usually evaluated and explored in the Casimir energy is the study of the massless limit of the field. In our study, the direct calculation of the massless limit from the above equation yields a divergent result. Thus, to obtain the finite answer for this particular limit, we return to Eq. (\ref{Delta.vac.DBC.1}) and let the mass of the field be zero. Thus, for the mass $m=0$, the APSF given in Eq. (\ref{APSF.INTEGER}) is applied to the summation expressions of Eq. (\ref{Delta.vac.DBC.1}) as follows:
\begin{eqnarray}\label{Delta.vac.DBC.Massless.2}
   \Delta \mathcal{E}_{\mbox{\tiny$\mathcal{D}$, vac}}^{(1)[m=0]}&=&\frac{-\lambda \pi^2\ell^{2(1-\xi)}}{8(2\pi)^6a}
      \Bigg\{\Big[\frac{-1}{2}f^{\mbox{\tiny[$0$]}}_{0}(a,\xi)\nonumber\\
     &&\hspace{-0.5cm}+\underbrace{\int_{0}^{\infty}f^{\mbox{\tiny[$0$]}}_x(a,\xi)dx}_{\mathcal{J}_1(a;\infty)}
     +B_1(0,a;\xi)\Big]^2
     \nonumber\\&&\hspace{-0.5cm}-\frac{1}{4}f^{\mbox{\tiny[$0$]}}_0(a,\xi)^2+\underbrace{\frac{1}{2}
     \int_{0}^{\infty}f^{\mbox{\tiny[$0$]}}_{x}(a,\xi)^2dx}_{\mathcal{J}_2(a;\infty)}
     \nonumber\\
     &&\hspace{-0.5cm}+\frac{1}{2}B_2(0,a;\xi)\Bigg\}+2\times\{a\to\frac{L-a}{2}\}\nonumber\\
     &&\hspace{-0.5cm}-\{a\to b\}-2\times\{a\to\frac{L-b}{2}\},
\end{eqnarray}
The use of APSF not only converts the summation expressions into the integral form, but also clarify the divergent parts of Eq. (\ref{Delta.vac.DBC.Massless.2}) like as: $\mathcal{J}^2_1(a;\infty)$, $\mathcal{J}_2(a;\infty)$, and the cross-term $\frac{1}{a}f^{\mbox{\tiny[$0$]}}_{0}(a,\xi)\mathcal{J}_1(a;\infty)$. To regularize and ultimately eliminate all of them, the scenario conducted in Eqs. (\ref{I.Disappear.1}), (\ref{I.Disappear.2}), and (\ref{cross.term.f_0DAR_I}) presuming $m=0$ should be reiterated. It causes all infinite contributions originating from these three terms to be removed.
Therefore, we obtain:
\begin{eqnarray}\label{Delta.vac.DBC.Massless.3}
       \Delta \mathcal{E}_{\mbox{\tiny$\mathcal{D}$, vac}}^{(1)[m=0]}&=&\frac{-\lambda \pi^2\ell^{2(1-\xi)}}{8(2\pi)^6a}\Big[2\mathcal{J}_1(a;\infty)B_1(0,a;\xi)\nonumber\\&-&
       f^{\mbox{\tiny[$0$]}}_0(a,\xi)B_1(0,a;\xi)+B_1(0,a;\xi)^2\nonumber\\&+&\frac{1}{2}B_2(0,a;\xi)\Big]
       +2\times\{a\to\frac{L-a}{2}\}
       \nonumber\\&-&\{a\to b\}-2\times\{a\to\frac{L-b}{2}\}.
\end{eqnarray}
The first two terms in the bracket of Eq. (\ref{Delta.vac.DBC.Massless.3}) are still divergent owing to the functions $f^{\mbox{\tiny[0]}}_0$ and $\mathcal{J}_1$ . To remove their infinities, we need to use the cutoff regularization technique. Hence, for the terms related to regions $\mathbf{a1}$, we substituted the upper limit of integrals $\mathcal{J}_1(a;\infty)$ and $f^{\mbox{\tiny[0]}}_0(a,\xi)$ for $\Omega_{\mathbf{a1}}$ and ${\Omega'}_{\mathbf{a1}}$, respectively. In the same way, the upper limit of integrals associated for each region $\mathbf{a2}$, $\mathbf{b1}$, and $\mathbf{b2}$ was determined. Then, by computing the integrals, we obtained:
\begin{eqnarray}\label{Delta.vac.DBC.1.Massless}
       \Delta \mathcal{E}_{\mbox{\tiny$\mathcal{D}$, vac}}^{(1)[m=0]}&=&\frac{-\lambda \pi^2\ell^{2(1-\xi)}}{8(2\pi)^6a}      \Big[\frac{4a\Omega_{\mathbf{a1}}^{3-\xi}}{3-\xi}B_1(0,a;\xi)\nonumber\\&-&
       \frac{2\pi{\Omega'}_{\mathbf{a1}}^{2-\xi}}{2-\xi}B_1(0,a;\xi)+B_1(0,a;\xi)^2\nonumber\\&+&\frac{1}{2}B_2(0,a;\xi)\Big]+2\times\{a\to\frac{L-a}{2}\}
       \nonumber\\&-&\{a\to b\}-2\times\{a\to\frac{L-b}{2}\}.
\end{eqnarray}
To remove divergences originating from cutoff values in Eq. (\ref{Delta.vac.DBC.1.Massless}), various types of adjustment for cutoffs $\Omega$ and ${\Omega'}$ are possible. Simply, we adjusted the cutoffs $\Omega$ and ${\Omega'}$ as:
\begin{eqnarray}\label{adjusting.cutoffs.Omega}
      \frac{\Omega_\mathbf{b1}^{3-\xi}}{\Omega_\mathbf{a1}^{3-\xi}}&=&\frac{B_1(0,a;\xi)}{B_1(0,b;\xi)},\hspace{0.9cm}
      \frac{\Omega_\mathbf{b2}^{3-\xi}}{\Omega_\mathbf{a2}^{3-\xi}}=\frac{B_1(0,\frac{L-a}{2};\xi)}{B_1(0,\frac{L-b}{2};\xi)},\nonumber\\
      \frac{{\Omega'}_\mathbf{b1}^{2-\xi}}{{\Omega'}_\mathbf{a1}^{2-\xi}}&=&\frac{bB_1(0,a;\xi)}{aB_1(0,b;\xi)},\nonumber\\
      \frac{{\Omega'}_\mathbf{b2}^{2-\xi}}{{\Omega'}_\mathbf{a2}^{2-\xi}}&=&\frac{(L-b)B_1(0,\frac{L-a}{2};\xi)}{(L-a)B_1(0,\frac{L-b}{2};\xi)}.
\end{eqnarray}
For any values of $\xi$, the above adjustments for the cutoffs cause all divergent parts related to $\Omega$s and ${\Omega'}$s to be removed via the BSS embedded in Eq. (\ref{Delta.vac.DBC.1.Massless}). Therefore, we obtain:
\begin{eqnarray}\label{DBC.Casimir.En.Massless}
     \Delta \mathcal{E}_{\mbox{\tiny $\mathcal{D}$, vac.}}^{\mbox{\tiny$(1)$}[m=0]}&=&\frac{-\lambda L^2\pi^2\ell^{2(1-\xi)}}{8(2\pi)^6a}      \Big[B_1(0,a;\xi)^2\nonumber\\&+&\frac{1}{2}B_2(0,a;\xi)\Big]+2\times\{a\to\frac{L-a}{2}\}
       \nonumber\\&-&\{a\to b\}-2\times\{a\to\frac{L-b}{2}\},
\end{eqnarray}
As Eq. (\ref{Calculation.B_2(a).No1}) shows, the value of the branch-cut term $B_2$ for any values of $\xi$ is exactly zero. Furthermore, the multiplying factor $\sin(\pi\xi/2)$ in Eq. (\ref{branch-cut.massless.1}) indicates that the branch-cut term $B_1(0,a;\xi)$ for any Even values of $\xi$ is exactly zero. Consequently, using Eq. (\ref{DBC.Casimir.En.Massless}), we can simply conclude that the RC to the Casimir energy in the massless case for any Even $\xi$ is zero. For any Odd values of $\xi$, Eq. (\ref{DBC.Casimir.En.Massless}) is converted to:
\begin{eqnarray}\label{DBC.Casimir.En.Massless.final}
       \mathcal{E}_{\mbox{\tiny $\mathcal{D},\xi$}}^{\mbox{\tiny$(1)$Cas.}[m=0]}&=&\frac{-\lambda \pi^2\ell^{2(1-\xi)}}{8(2\pi)^6a}\Big[a^{2\xi-4}\tilde{\mathcal{B}}^2_\xi(\epsilon_\mathbf{a1})\Big]\nonumber\\&+&2\times\{a\to\frac{L-a}{2}\}
       -\{a\to b\}\nonumber\\&-&2\times\{a\to\frac{L-b}{2}\},
\end{eqnarray}
For $\xi=1$, using Eqs. (\ref{BSS.Definition}), (\ref{DBC.Casimir.En.Massless.final}), and (\ref{Calculation.B_1(a).No2.2}), the Casimir energy density is obtained as:
\begin{eqnarray}\label{massless.Cas.energy.xi=1}
       \mathcal{E}_{\mbox{\tiny $\mathcal{D},1$}}^{\mbox{\tiny$(1)$Cas.}[m=0]}=\frac{-\lambda}{18432a^3}
\end{eqnarray}
This result is completely authenticated with the previously reported result for the system without a Lifshitz parameter\,\cite{RC.6}. For all Odd values of $\xi\geq3$, the function $\tilde{\mathcal{B}}_\xi(\epsilon)$ is divergent at the limit $\epsilon\to0$. To remove their divergences from Eq. (\ref{DBC.Casimir.En.Massless.final}), proper adjustings for regulators $\epsilon_\mathbf{a1}$, $\epsilon_\mathbf{a2}$, $\epsilon_\mathbf{b1}$, $\epsilon_\mathbf{b2}$, $L$, and $b$ should be performed. For instance, in the case of $\xi=3$, by adjusting regulators as $\epsilon_\mathbf{a1}=\epsilon_\mathbf{b1}=\epsilon_\mathbf{a2}=\epsilon_\mathbf{b2}=\epsilon$, Eq. (\ref{DBC.Casimir.En.Massless.final}) is converted to:
\begin{eqnarray}\label{massless.Cas.energy.xi=3}
        \mathcal{E}_{\mbox{\tiny $\mathcal{D},3$}}^{\mbox{\tiny$(1)$Cas.}[m=0]}&=&\frac{-\lambda \pi^2}{8\ell^4(2\pi)^6}\\&&\times\Big[a+2\frac{L-a}{2}-b-2\frac{L-b}{2}\Big]\tilde{\mathcal{B}}^2_3(\epsilon)=0.\nonumber
\end{eqnarray}
Eq. (\ref{massless.Cas.energy.xi=3}) demonstrates that the RC to the Casimir energy of the massless Lifshitz - like scalar field for $\xi=3$ is exactly equal to zero. The same result is available for all Odd values of $\xi>3$. In fact, proper adjustment for regulators $\epsilon_\mathbf{a1}$, $\epsilon_\mathbf{a2}$, $\epsilon_\mathbf{b1}$, $\epsilon_\mathbf{b2}$, $L$, and $b$ provides a situation in which all divergences are removed from Eq. (\ref{DBC.Casimir.En.Massless.final}). Due to the participation of a significant number of regulators in Eq. (\ref{DBC.Casimir.En.Massless.final}), there are always enough degrees of freedom for the proper adjustment of regulators. Consequently, we can infer that the Casimir energy for the massless Lifshitz-like scalar field leads to zero for any $\xi\geq3$.
\par
To obtain the RC to the Casimir energy for the self-interacting Lifshitz-like scalar field confined between two parallel plates with NBC/PBC, it is required all the computation process, analogous to what conducted for DBC, be performed. We do not intend to reiterate their computations here. Only from the cognitions of the nature of Periodic and Neumann boundary conditions and their relations with the Dirichlet one, we can simply obtain the following forms for the Casimir energy\,\cite{wolfram,INJP}:
\begin{eqnarray}\label{relation of.BCs}
      \mathcal{E}_{\mbox{\tiny $\mathcal{N},\xi$}}^{\mbox{\tiny$(1)$Cas.}}(a)&=&\mathcal{E}_{\mbox{\tiny $\mathcal{D},\xi$}}^{\mbox{\tiny$(1)$Cas.}}(a),\nonumber\\ \mathcal{E}_{\mbox{\tiny $\mathcal{P},\xi$}}^{\mbox{\tiny$(1)$Cas.}}(a)&=&2\mathcal{E}_{\mbox{\tiny $\mathcal{D},\xi$}}^{\mbox{\tiny$(1)$Cas.}}(a/2)
\end{eqnarray}
where $\mathcal{E}_{\mbox{\tiny $\mathcal{P},\xi$}}^{\mbox{\tiny$(1)$Cas.}}(a)$, $\mathcal{E}_{\mbox{\tiny $\mathcal{D},\xi$}}^{\mbox{\tiny$(1)$Cas.}}(a)$ and $\mathcal{E}_{\mbox{\tiny $\mathcal{N},\xi$}}^{\mbox{\tiny$(1)$Cas.}}(a)$ denote the Casimir energy density (per unit area) for Periodic, Dirichlet and
Neumann boundary conditions between a pair of plates with the distance $a$, respectively. The relations displayed in Eq. (\ref{relation of.BCs}) for both cases of massive and massless scalar fields are valid \cite{3D-Reza}.
\begin{figure}[th]\centering\includegraphics[width=8cm]{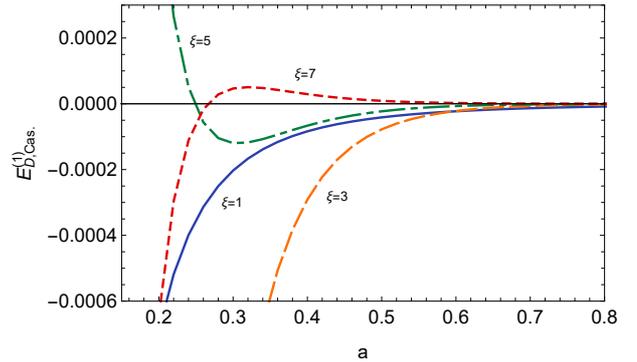}
\caption{\label{aPlotDBC}
The plot of the first-order RC to the Casimir energy density (per unit area) of the massive Lifshitz-like scalar field between a pair of plates with the distance $a$ for $m=1$ and $\lambda=0.1$; the figure also presents the sequence of plots for the values of critical exponents $\xi=\{1,3,5,7\}$. All units, including the distance of the plates ($a$) as well as the Casimir energy, are considered in the natural unit ($\hbar c=1$).}
\end{figure}
\begin{figure}[th]\centering\includegraphics[width=8cm]{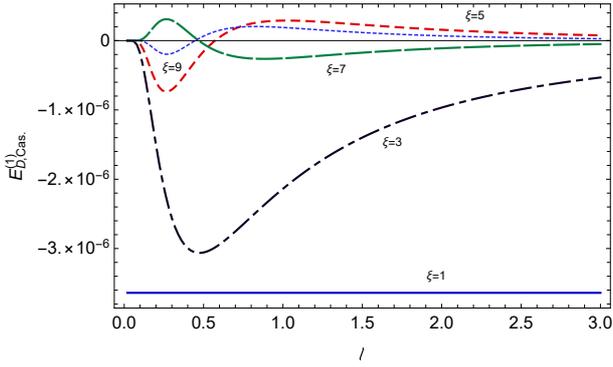}
\caption{\label{ELLPLOTDBC}
The first-order RC to the Casimir energy density (per unit area) for a special distance of plates ($a=1$) within the self-interacting massive Lifshitz-like scalar field was plotted as a function of $\ell$. In this figure, we have shown the sequence of plots for critical exponents $\xi=\{1,3,5,7,9\}$. The values of the mass and coupling constant in all plots are considered as $m=1$ and $\lambda=0.1$, respectively. All units, including the mass of the field, the
distance of the plates $a$, and the Casimir energy, were considered in the natural unit ($\hbar c=1$).}
\end{figure}
\begin{figure}[th]\centering\includegraphics[width=8cm]{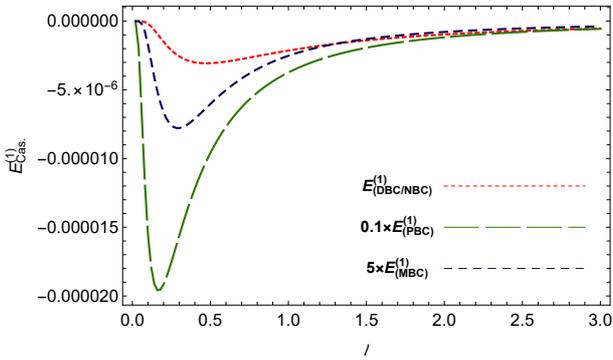}
\caption{\label{ELLDNPM-BC}
The first-order RC to the Casimir energy per unit area for the massive Lifshitz-like scalar field confined between two parallel plates by the distance $a=1$ with four Dirichlet, Neumann, Periodic, and Mixed boundary conditions as a function of $\ell$ is plotted. The subscripts DBC, NBC, PBC and MBC denote the type of the boundary condition, namely Dirichlet, Neumann, Periodic, and Mixed boundary conditions, respectively. The values of the mass and coupling constant in all plots are considered $m=1$ and $\lambda=0.1$, respectively. In addition, in all plots, the critical exponent is $\xi=3$. All units, including the mass of the field, the distance of the plates $a$, and the Casimir energy, are considered in the natural unit ($\hbar c=1$).}
\end{figure}
\subsection{Mixed Boundary Condition}\label{subsec: MBC.RC}
To obtain the RC to the Casimir energy for the massive Lifshitz-like scalar field confined with MBC between two parallel plates according to the definition form of the Casimir energy introduced in Eq. (\ref{BSS.Definition}), the vacuum energy of each region in Fig. (\ref{fig.1}) should be at hand. Hence, we put the distance of the plates corresponding to each region shown in Fig. (\ref{fig.1}), into Eq. (\ref{Green's.Func.Mixed}). Then, using Eqs. (\ref{BSS.Definition}) and (\ref{vacuum.energy.EXP2}), we subtract the vacuum energy densities for two displayed configurations of Fig. (\ref{fig.1}) as follows:
\begin{eqnarray}\label{Delta.vac.MBC.1}
     \Delta \mathcal{E}_{\mbox{\tiny$\mathcal{M}$, vac}}^{(1)[m]}&=&\frac{-\lambda \pi^2\ell^{2(1-\xi)}}{4(2\pi)^6a}\nonumber\\&\times&\Bigg\{\sum_{n,n'=0}^{\infty}
f^{\mbox{\tiny[$m$]}}_{n+\frac{1}{2}}(a,\xi)f^{\mbox{\tiny[$m$]}}_{n'+\frac{1}{2}}(a,\xi)\nonumber\\
     &+&\frac{1}{2}\sum_{n=0}^{\infty}\Big[f^{\mbox{\tiny[$m$]}}_{n+\frac{1}{2}}(a,\xi)\Big]^2\Bigg\}+2\times\{a\to\frac{L-a}{2}\}
\nonumber\\&-&\{a\to b\}-2\times\{a\to\frac{L-b}{2}\}.
\end{eqnarray}
When the expressions of Eq. (\ref{Delta.vac.MBC.1}) are in the summation form, the removal procedure of their infinities is cumbersome. Hence, it better to change the current form of divergences to the integral form. Therefore, to convert the summation forms of Eq. (\ref{Delta.vac.MBC.1}) into the integral form, the following phrase for APSF was used \cite{A.Saharian},
\begin{eqnarray}\label{APSF.Half.INTEGER}
      \sum_{n=0}^{\infty}\mathcal{F}(n+\frac{1}{2})&=&\int_{0}^{\infty}\mathcal{F}(x)dx
      -i\int_{0}^{\infty}\frac{\mathcal{F}(it)-\mathcal{F}(-it)}{e^{2\pi t}+1}dt.\nonumber\\
\end{eqnarray}
Applying this form of APSF on all summations of Eq. (\ref{Delta.vac.MBC.1}) converts it as follows:
\begin{eqnarray}\label{Delta.vac.MBC.after.APSF}
       \Delta \mathcal{E}_{\mbox{\tiny$\mathcal{M}$, vac}}^{(1)[m]}&=&\frac{-\lambda \pi^2\ell^{2(1-\xi)}}{4(2\pi)^6a}
      \nonumber\\&\times&\Bigg\{\Big[\underbrace{\int_{0}^{\infty}f^{\mbox{\tiny[$m$]}}_x(a,\xi)dx}_{\mathcal{I}_1(a;\infty)}
      +B_1(m,a;\xi)\Big]^2
      \nonumber\\&+&\underbrace{\frac{1}{2}\int_{0}^{\infty}f^{\mbox{\tiny[$m$]}}_{x}(a,\xi)^2dx}_{\mathcal{I}_2(a;\infty)}
     +\frac{1}{2}B_2(m,a;\xi)\Bigg\}\nonumber\\&+&2\times\{a\to\frac{L-a}{2}\}-\{a\to b\}\nonumber\\
      &&\hspace{-0.5cm}-2\times\{a\to\frac{L-b}{2}\}.
\end{eqnarray}
The functions $\mathcal{I}_1(a;\infty)$ and $\mathcal{I}_2(a;\infty)$ and all phrases that are a multiple of them, denoted in Eq. (\ref{Delta.vac.MBC.after.APSF}), are divergent. To remove their divergences, the scenario considered in the previous subsection is followed. Therefore, analogous to what conducted in Eqs. (\ref{I.Disappear.1}) and (\ref{I.Disappear.2}), we eliminated the contribution of terms $\mathcal{I}^2_1$ and $\mathcal{I}_2$ from Eq. (\ref{Delta.vac.MBC.after.APSF}). It converts Eq. (\ref{Delta.vac.MBC.after.APSF}) to:
\begin{eqnarray}\label{Delta.vac.MBC.after.APSF.2}
       \Delta \mathcal{E}_{\mbox{\tiny$\mathcal{M}$, vac}}^{(1)[m]}&=&\frac{-\lambda \pi^2\ell^{2(1-\xi)}}{4(2\pi)^6a}
      \nonumber\\&\times&\Big[
      B_1(m,a;\xi)^2+2\mathcal{I}_1(a;\infty)B_1(m,a;\xi)\nonumber\\&+&\frac{1}{2}B_2(m,a;\xi)\Big]+2\times\{a\to\frac{L-a}{2}\}
      \nonumber\\&&\hspace{-0.5cm}-\{a\to b\}-2\times\{a\to\frac{L-b}{2}\}.
\end{eqnarray}
For $\xi\leq3$, the second term in Eq. (\ref{Delta.vac.MBC.after.APSF.2}) is still divergent owing to the upper limit of the integral $\mathcal{I}_1$. The process of eliminating the divergences resulted from the integral $\mathcal{I}_1$, and finding its convergent contribution is similar to what was followed in Eqs. (\ref{I_1.Value.for.xi>3}) to (\ref{Finite.remained.I_1(a)}). Therefore, using the results presented in Eq. (\ref{Finite.remained.I_1(a)}), we can rewrite Eq. (\ref{Delta.vac.MBC.after.APSF.2}) as follows:
\begin{eqnarray}\label{Delta.vac.MBC.after.APSF.3}
       \Delta \mathcal{E}_{\mbox{\tiny$\mathcal{M}$, vac}}^{(1)[m]}&=&\frac{-\lambda \pi^2\ell^{2(1-\xi)}}{4(2\pi)^6a}
      \nonumber\\&\times&\Big[
      B_1(m,a;\xi)^2+2\tilde{\mathcal{I}}(a)B_1(m,a;\xi)\nonumber\\&+&\frac{1}{2}B_2(m,a;\xi)\Big]+2\times\{a\to\frac{L-a}{2}\}
      \nonumber\\&&\hspace{-0.5cm}-\{a\to b\}-2\times\{a\to\frac{L-b}{2}\}.
\end{eqnarray}
Now, the limits $L,b\to\infty$, expressed in Eq. (\ref{BSS.Definition}), should be applied. The convergence behavior of the branch-cut terms in large distance of plates, has caused all finite contributions remained from the vacuum energy of regions $\mathbf{a2}$, $\mathbf{b1}$ and $\mathbf{b2}$ to be vanished. In addition, using Eq. (\ref{Calculation.B_2(a).No1}), the branch-cut term $B_2(m,a;\xi)$ is zero for any values of $\xi$. Consequently, the one-Loop RC to the Casimir energy density for the massive Lifshitz-Like scalar field confined by MBC between a pair of plates with the distance $a$ is obtained as:
\begin{eqnarray}\label{RC.Cas.MBC.Final.Massive}
        \mathcal{E}_{\mbox{\tiny $\mathcal{M},\xi$}}^{\mbox{\tiny$(1)$Cas.}[m]}&=&\frac{-\lambda \pi^2\ell^{2(1-\xi)}}{4(2\pi)^6a}
      \nonumber\\&\times&B_1(m,a;\xi)\Big[
      B_1(m,a;\xi)+2\tilde{\mathcal{I}}_1(a)\Big].
\end{eqnarray}
\par
To calculate the massless limit for the above result, by presuming $m=0$, we return to Eq. (\ref{Delta.vac.DBC.1}) and reiterate the calculations. Hence, similar to what performed in the massive case, and to convert the summation forms in Eq. (\ref{Delta.vac.MBC.1}) into the integral form, we again used the APSF given in Eq. (\ref{APSF.Half.INTEGER}). Thus, we obtain:
\begin{eqnarray}\label{Delta.vac.massless.MBC.1}
         \Delta \mathcal{E}_{\mbox{\tiny$\mathcal{M}$, vac}}^{(1)[m=0]}&=&\frac{-\lambda \pi^2\ell^{2(1-\xi)}}{4(2\pi)^6a}
\Bigg[\mathcal{J}_1(a;\infty)^2+B_1(0,a;\xi)^2\nonumber\\ &&\hspace{-0.5cm}+2\mathcal{J}_1(a;\infty)B_1(0,a;\xi)+\mathcal{J}_2(a;\infty)\nonumber\\&&\hspace{-0.5cm}
+\frac{1}{2}B_2(0,a;\xi)\Bigg]+2\times\{a\to\frac{L-a}{2}\}\nonumber\\
     &&\hspace{-0.5cm}-\{a\to b\}-2\times\{a\to\frac{L-b}{2}\},
\end{eqnarray}
where
\begin{eqnarray}\label{Defining.J1&J2}
   \mathcal{J}_1(a;\infty)&=&\int_{0}^{\infty}f^{\mbox{\tiny[$0$]}}_x(a,\xi)dx\nonumber\\ \mathcal{J}_2(a;\infty)&=&\frac{1}{2}\int_{0}^{\infty}f^{\mbox{\tiny[$0$]}}_{x}(a,\xi)^2dx.
\end{eqnarray}
In Eq. (\ref{Delta.vac.massless.MBC.1}), two terms $\mathcal{J}^2_1(a;\infty)$ and $\mathcal{J}_2(a;\infty)$ are divergent. To regularize and ultimately eliminate both of them, the scenario conducted in Eqs. (\ref{I.Disappear.1}) and (\ref{I.Disappear.2}) presuming $m=0$ should be reiterated. It causes all infinite contributions originating from these two terms to be removed. Then, it is the turn of the cross-term $\mathcal{J}_1\times B_1$. This term is also divergent owing to the upper limit of integral $\mathcal{J}_1$. To regularize its infinity, the upper limit of the integral $\mathcal{J}_1$ was replaced with a cutoff parameter as done in $\Lambda_\mathbf{a1}$. Similarly, the cutoffs $\Lambda_\mathbf{a2}$, $\Lambda_\mathbf{b1}$, and $\Lambda_\mathbf{b2}$ were replaced on the upper limit of $\mathcal{J}_1$ related to the other regions.  Therefore, Eq. ({\ref{Delta.vac.massless.MBC.1}}) was converted to:
\begin{eqnarray}\label{Delta.vac.massless.MBC.2}
         \Delta \mathcal{E}_{\mbox{\tiny$\mathcal{M}$, vac}}^{(1)[m=0]}&=&\frac{-\lambda \pi^2\ell^{2(1-\xi)}}{4(2\pi)^6a}
      \nonumber\\ \hspace{-0.5cm}&\times&
\Big[\frac{4a\Lambda_\mathbf{a1}^{3-\xi}}{3-\xi}B_1(0,a;\xi)+B_1(0,a;\xi)^2\nonumber\\&&\hspace{-0.5cm}
     +\frac{1}{2}B_2(0,a;\xi)\Big]+2\times\{a\to\frac{L-a}{2}\}\nonumber\\
     &&\hspace{-0.5cm}-\{a\to b\}-2\times\{a\to\frac{L-b}{2}\}.
\end{eqnarray}
To remove the divergences originating from the first term in the bracket of the above equation, various methods are feasible to adjust the cutoff $\Lambda$ are feasible. Simply, we adjust the cutoff as:
\begin{eqnarray}\label{adjusting.cutoffs.Lambda.MBC.massless}
      \frac{\Lambda_\mathbf{b1}^{3-\xi}}{\Lambda_\mathbf{a1}^{3-\xi}}=\frac{B_1(0,a;\xi)}{B_1(0,b;\xi)}\hspace{0.9cm}
      \frac{\Lambda_\mathbf{b2}^{3-\xi}}{\Lambda_\mathbf{a2}^{3-\xi}}=\frac{B_1(0,\frac{L-a}{2};\xi)}{B_1(0,\frac{L-b}{2};\xi)}.
\end{eqnarray}
For any values of $\xi$, the above adjustments for the cutoff cause all divergent parts related to $\Lambda$s to be removed via the BSS embedded in Eq. (\ref{Delta.vac.massless.MBC.2}). Now, based on Eq. (\ref{Calculation.B_2(a).No1}) which acclaim the branch-cut term $B_2(0,a;\xi)$ as zero, we obtain:
\begin{eqnarray}\label{MBC.Casimir.En.Massless.final}
      \Delta \mathcal{E}_{\mbox{\tiny $\mathcal{M},\xi$}}^{\mbox{\tiny(1)Cas.[m=0]}}&=&\frac{-\lambda \pi^2\ell^{2(1-\xi)}}{4(2\pi)^6}\Big[\frac{B_1(0,a;\xi)^2}{a}\Big]
      \nonumber\\ \hspace{-0.5cm}
&+&2\times\{a\to\frac{L-a}{2}\}-\{a\to b\}\nonumber\\
     &&\hspace{-0.5cm}-2\times\{a\to\frac{L-b}{2}\}.
\end{eqnarray}
The multiplying factor $\sin(\pi\xi/2)$ in Eq. (\ref{branch-cut.massless.1}) shows that the branch-cut term $B_1(0,a;\xi)$ for any Even values of $\xi$ is exactly zero. Consequently, we can simply conclude that the RC to the Casimir energy in the massless case for any Even $\xi$ is zero. For Odd values of $\xi$, using Eq. (\ref{MBC.Casimir.En.Massless.final}), the above Equation is converted to:
\begin{eqnarray}\label{ECas.massless.MBC.Odd.xi}
       \Delta \mathcal{E}_{\mbox{\tiny $\mathcal{M},\xi$}}^{\mbox{\tiny(1)Cas.[m=0]}}&=&\frac{-\lambda \pi^2\ell^{2(1-\xi)}}{4(2\pi)^6}\Big[a^{2\xi-5}\tilde{\mathcal{B}}_\xi(\epsilon_\mathbf{a1})^2\Big]
      \nonumber\\ \hspace{-0.5cm}
&+&2\times\{a\to\frac{L-a}{2}\}-\{a\to b\}\nonumber\\
     &&\hspace{-0.5cm}-2\times\{a\to\frac{L-b}{2}\}.
\end{eqnarray}
This equation for all Odd values of $\xi\neq1$ is divergent. A simple comparison between Eqs. (\ref{DBC.Casimir.En.Massless.final}) and (\ref{ECas.massless.MBC.Odd.xi}) indicates that the type of divergence that appeared in their is the same. This similarity allows the technique, which is used to eliminate the divergences in Eq. (\ref{DBC.Casimir.En.Massless.final}), to be employed here. Hence, adjustment of the proper value for regulators, such as $\epsilon_\mathbf{a1}$, $\epsilon_\mathbf{a2}$, $\epsilon_\mathbf{b1}$, $\epsilon_\mathbf{b2}$, $L$, and $b$ in Eq. (\ref{ECas.massless.MBC.Odd.xi}), along with the BSS embedded in Eq. (\ref{ECas.massless.MBC.Odd.xi}), will remove all divergent contributions made by the function $\tilde{\mathcal{B}}_\xi(\epsilon)$. Consequently, we can now declare that the Casimir energy for the massless Lifshitz-like scalar field is zero for any $\xi\neq1$. The only remained case is $\xi=1$, that is corresponding to the scalar field without a Lifshitz critical exponent. The RC to the Casimir energy density for this case using Eqs. (\ref{Calculation.B_1(a).No2.2}) is obtained as:
\begin{eqnarray}\label{MBC.xi=1.massless}
\Delta \mathcal{E}_{\mbox{\tiny $\mathcal{M},\xi$}}^{\mbox{\tiny(1)Cas.[m=0]}}=\frac{-\lambda}{36864a^3},
\end{eqnarray}
which is perfectly consistent with the previously reported result without a Lifshitz critical exponent \cite{mixed.}. The first order RC to the Casimir energy for the massive Lifshitz-like scalar field confined with a pair of plates with DBC was plotted in Fig. (\ref{aPlotDBC}). In this plot, the Casimir energy value per unit area as a function of the distance of the plates ($a$) for different values of $\xi$ was plotted. This plot shows that for $\xi\leq3$, RC to the Casimir energy is negative and consequently its related force is attractive. This behavior does not hold for $\xi>3$, since the sign of the RC to the Casimir energy has alteration for $\xi>3$. In  Fig. (\ref{ELLPLOTDBC}), to find the Casimir energy behavior as a function of $\ell$ and the Lifshitz critical exponent $\xi$, the Casimir energy for the specific value of the mass of the field ($m=1$), coupling constant $\lambda=0.1$ and distance of plates $a=1$ were plotted. The sequence of plots in this figure shows a maximum alteration in the Casimir energy for each $\xi$, and the value of this extremum in the Casimir energy decreases by increasing $\xi$. In Fig. (\ref{ELLDNPM-BC}), the orders of the Casimir energy density for different boundary conditions in the specific values of $a$ and $m$ are compared to each other. This plot is useful in manifesting the influence of the Lifshitz symmetry breaking on RC to the Casimir energy for different boundary conditions.
\section{Concluding Remarks}\label{sec:conclusion}
In this paper, RC to the Casimir energy for the Lifshitz-like scalar field confined between DBC/MBC was computed. This computation for both massive and massless cases were performed. Moreover, using the results obtained for the case of DBC, and by identifying the nature of NBC and PBC, the Casimir energy for these boundary conditions was generalized. In this calculation, a systematic method was used to renormalize the bare parameters of Lagrangian. Our used procedure allows all influences from the boundary conditions to be imported in the renormalization program. One of the main results of importing the influences of the boundary condition in the renormalization program is the appearance of position-dependent counterterms. This renormalization program, along with BSS as a regularization technique has constructed a clear and unambiguous way to calculate the Casimir energy. Using this computation process, for any values of the critical exponent $\xi$, RC to the Casimir energy of the massive Lifshitz-like scalar field is finite and satisfies physical expectations. Furthermore, RC to the Casimir energy for all the massless Lifshitz-like scalar field (except for $\xi=1$) was obtained as zero. This quantity for the massive / massless case is consistent with the system without a Lifshitz critical exponent ($\xi=1$).
\appendix
\section{Calculation of The Green's Function}\label{Appendix.Green.Function}
\setcounter{equation}{0}
\renewcommand{\theequation}{\Alph{section}.\arabic{equation}}
To obtain the Green's function expression, we start with the following differential equation:
\begin{eqnarray}\label{dalambrian.green.func.}
      \Big[\Box+m^2\Big]G(x,x')=-\delta(x-x'),
\end{eqnarray}
where $\Box=\partial_0^2+\ell^{2(\xi-1)}(-1)^{\xi}(\partial^2_x+\partial^2_y+\partial^2_z)^{\xi}$. We set the Green's function form as:
\begin{eqnarray}\label{first.green.func.form}
     G(x,x')=\sum_{n=1}^{\infty}\int\frac{d\omega}{2\pi}\int\frac{d^2\bf{k}}{(2\pi)^2}\mathcal{A}(x')\phi_n(x),
\end{eqnarray}
where $\phi_n(x)=e^{i\omega t}e^{i\mathbf{k}\cdot\mathbf{X}}\sin[\frac{n\pi}{a}(z+\frac{a}{2})]$ is the eigenfunction obtained from Eqs. (\ref{Klein-Gordon.Eq}) and (\ref{Dirichlet.BC.}). We can also define the delta function $\delta(x-x')$ as:
\begin{eqnarray}\label{delta.function.form}
      \delta(x-x')=\frac{2}{a}\sum_{n=1}^{\infty}\int\frac{d\omega}{2\pi}\int\frac{d^2\bf{k}}{(2\pi)^2}
      \phi_n(x)\phi^{\ast}_n(x')
\end{eqnarray}
where $\mathbf{k}=(k_x,k_y)$. Now, using Eq. (\ref{delta.function.form}) and substituting the prescribed form of the Green's function given in Eq. (\ref{first.green.func.form}) for Eq. (\ref{dalambrian.green.func.}), we obtain:
\begin{eqnarray}\label{coefficient}
      \mathcal{A}(x')=\frac{2}{a}\frac{\phi^{\ast}_n(x')}
      {-\omega^2+\omega^{\mbox{\tiny($\mathcal{D}$)}}_{\xi,n}(\mathbf{k})^2}
\end{eqnarray}
where $\omega^{\mbox{\tiny(D)}}_{\xi,n}(\mathbf{k})
=\big[\ell^{2(\xi-1)}\big(\mathbf{k}^2+\big(\frac{n\pi}{a}\big)^2\big)^{\xi}+m^2\big]^{1/2}$. Substituting $\mathcal{A}(x')$ for Eq. (\ref{first.green.func.form}) provides the Green's function expression as:
\begin{eqnarray}\label{green.func.form.before.wick.rotation}
      G(x,x')&=&\frac{2}{a}\int\frac{d\omega}{2\pi}\int\frac{d^2\mathbf{k}}{(2\pi)^2}\nonumber\\&&\times\sum_{n=1}^{\infty}\frac{\begin{array}{c}
                                                                      e^{i\omega(t-t')}e^{i\mathbf{k}(\mathbf{x}-\mathbf{x}')}\\
                                                                      \times\sin[\frac{n\pi}{a}(z+\frac{a}{2})]\sin[\frac{n\pi}{a}(z'+\frac{a}{2})]
                                                                    \end{array}}{-\omega^2+\omega^{\mbox{\tiny$(\mathcal{D})$}}_{\xi,n}(\mathbf{k})^2},\nonumber\\
\end{eqnarray}
Eventually, applying the Wick rotation trick and changing the variable $k=(i\omega,\mathbf{k})$ are the final steps converting the above expression to the form of Green's function presented in Eq. (\ref{Green's.Func.Dirichlet}). Moreover, performing the same steps for the problem in the mixed boundary condition will result in the Green's function expression presented in Eq. (\ref{Green's.Func.Mixed}).

\section{Calculation of Branch-Cut Terms}\label{Appendix.1}
\setcounter{equation}{0}
\renewcommand{\theequation}{\Alph{section}.\arabic{equation}}
The last term on the right hand side of Eqs. (\ref{APSF.INTEGER}) and (\ref{APSF.Half.INTEGER}) is usually known as the branch-cut term. Using its definition we can obtain $B_1(m,a;\xi)$ as follows:
\begin{eqnarray}\label{Calculation.B_1(a).No1}
           B_1(m,a;\xi)=2\pi ic\int_{0}^{\infty}dt\int_{0}^{\infty}kdk\frac{\mathcal{H}_{\xi}(a)-\mathcal{H}^{\ast}_{\xi}(a)}{e^{2\pi t}-c}
\end{eqnarray}
where $\mathcal{H}_{\xi}(a)=e^{-i\pi\xi/2}[(-k^2+(\frac{t\pi}{a})^2)^\xi+\mu^{2\xi}e^{-i\pi\xi}]^{-1/2}$ and $c=\pm1$. To obtain the branch-cut term $B_1(m,a;\xi)$ in the case of scalar field confined with DBC (Eq. (\ref{APSF.INTEGER})), the value of the parameter $c$ should be considered $c=1$, and the value $c=-1$ refers to the case of the problem in which the scalar field is confined with MBC (Eq. (\ref{APSF.Half.INTEGER})). For any values of $\xi$, after changing the order $k$ and $t$ in the integration, we obtain:
\begin{eqnarray}\label{Calculation.B_1(a).No2}
           B_1(m,a;\xi)&=&4ac\sin\Big(\frac{\pi\xi}{2}\Big)\int_{\mu}^{\infty}\frac{dT}{e^{2aT}-c}\nonumber\\ &&\times\int_{0}^{\sqrt{T^2-\mu^2}}\frac{kdk}{\sqrt{(-k^2+T^2)^\xi-\mu^{2\xi}}},
\end{eqnarray}
where $T=t\pi/a$. Computing the integral over the parameter $k$ leads to:
\begin{eqnarray}\label{computing the integral B1 over k}
          B_1(m,a;1)&=&2ac\sqrt{-\mu ^{-2\xi}}\sin\Big(\frac{\pi\xi}{2}\Big)\nonumber\\&\times&
          \int_{\mu}^{\infty}\frac{dT}{e^{2aT}-c}\Big[T^2\,
          _2F_1\Big(\frac{1}{2},\frac{1}{\xi};1+\frac{1}{\xi};\frac{T^{2\xi}}{\mu^{2\xi }}\Big)\nonumber\\&&\hspace{2.2cm}-\mu^2\,_2F_1\Big(\frac{1}{2},\frac{1}{\xi };1+\frac{1}{\xi };1\Big)\Big],\nonumber\\
\end{eqnarray}
where $_2F_1(\alpha_1,\alpha_2;\alpha_3;z)$ is the Hypergeometric function. As Eq. (\ref{computing the integral B1 over k}) displays, the value of the branch-cut term $B_1$ for all even values of $\xi$ is zero. For $\xi=1$, the above integration is converted to,
\begin{eqnarray}\label{Calculation.B_1(a).No2.2}
           B_1(m,a;1)&=&4ac\int_{\mu}^{\infty}\frac{\sqrt{T^2-\mu^2}}{e^{2aT}-c}dT=2\mu\sum_{j=1}^{\infty}\frac{c^j K_1(2\mu aj)}{j}\nonumber\\
\end{eqnarray}
where $K_1(\alpha)$ is the modified Bessel function. For other odd values of $\xi$, finding a closed form answer from the integration expressed in Eq. (\ref{computing the integral B1 over k}) is a cumbersome task. Hence, for odd values $\xi>1$, the computation of the branch-cut term $B_1$ was followed numerically.
For the massless scalar field, according to Eq. (\ref{Calculation.B_1(a).No1}), the branch-cut term $B_1(0,a;\xi)$ is converted to:
\begin{eqnarray}\label{branch-cut.massless.1}
      B_1(0,a;\xi)&=&4ca^{\xi-2}\sin\Big(\frac{\pi\xi}{2}\Big)
\nonumber\\&&\hspace{1cm}\times\int_{\epsilon\to0}^{\infty}d\eta\int_{0}^{\eta}\chi d\chi\frac{[\eta^2-\chi^2]^{-\xi/2}}{e^{2\eta}-c}\nonumber\\&&\hspace{1cm}=a^{\xi-2}\tilde{\mathcal{B}}_\xi(\epsilon),
\end{eqnarray}
where variables $T=\eta/a$ and $\chi=ka$ were changed. The above integration has a divergent value, particularly owing to the lower limit of integration over $\eta$. Therefore, to regularize its infinity, we replaced the lower limit of integration over $\eta$ with a regulator like as $\epsilon$. Substituting this form of the branch-cut term $B_1(0,a,\xi)$ in Eq. (\ref{DBC.Casimir.En.Massless}) and using the BSS supplemented by the cutoff regularization technique yield no contribution remained from $B_1(0,a,\xi)$ in the Casimir energy.
\par
For the branch-cut term $B_2(m,a;\xi)$, we have:
\begin{eqnarray}\label{Calculation.B_2(a).No1}
          B_2(m,a;\xi)&=&ic\int_{0}^{\infty}\frac{dt}{e^{2\pi t}-c}\\&&\hspace{-0.5cm}\times\Bigg[\left(\int d^2k \mathcal{H}_\xi(a)\right)^2-\left(\int d^2k \mathcal{H}^{\ast}_\xi(a)\right)^2\Bigg]=0\nonumber
\end{eqnarray}
This equation shows that for any values of $m$ and $\xi$, the branch-cut term $B_2(m,a;\xi)$ is exactly zero.
\begin{acknowledgements}
The author would like to thank the research office of Semnan Branch, Islamic Azad University for the financial support.
\end{acknowledgements}

\end{document}